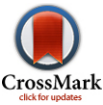

# Lamellar Thickness and Stretching Temperature Dependency of Cavitation in Semicrystalline Polymers

**Yaotao Wang, Zhiyong Jiang, Lianlian Fu, Ying Lu, Yongfeng Men\***

State Key Laboratory of Polymer Physics and Chemistry, Changchun Institute of Applied Chemistry, Chinese Academy of Sciences, University of Chinese Academy of Sciences, Changchun, People's Republic of China

## Abstract

Polybutene-1 (PB-1), a typical semicrystalline polymer, in its stable form I shows a peculiar temperature dependent strain-whitening behavior when being stretched at temperatures in between room temperature and melting temperature of the crystallites where the extent of strain-whitening weakens with the increasing of stretching temperature reaching a minima value followed by an increase at higher stretching temperatures. Correspondingly, a stronger strain-hardening phenomenon was observed at higher temperatures. The strain-whitening phenomenon in semicrystalline polymers has its origin of cavitation process during stretching. In this work, the effect of crystalline lamellar thickness and stretching temperature on the cavitation process in PB-1 has been investigated by means of combined synchrotron ultrasmall-angle and wide-angle X-ray scattering techniques. Three modes of cavitation during the stretching process can be identified, namely "no cavitation" for the quenched sample with the thinnest lamellae where only shear yielding occurred, "cavitation with reorientation" for the samples stretched at lower temperatures and samples with thicker lamellae, and "cavitation without reorientation" for samples with thinner lamellae stretched at higher temperatures. The mode "cavitation with reorientation" occurs before yield point where the plate-like cavities start to be generated within the lamellar stacks with normal perpendicular to the stretching direction due to the blocky substructure of the crystalline lamellae and reorient gradually to the stretching direction after strain-hardening. The mode of "cavitation without reorientation" appears after yield point where ellipsoidal shaped cavities are generated in those lamellae stacks with normal parallel to the stretching direction followed by an improvement of their orientation at larger strains. X-ray diffraction results reveal a much improved crystalline orientation for samples with thinner lamellae stretched at higher temperatures. The observed behavior of microscopic structural evolution in PB-1 stretched at different temperatures explains above mentioned changes in macroscopic strain-whitening phenomenon with increasing in stretching temperature and stress-strain curves.

Funding: This work is supported by the National Natural Science Foundation of China (21134006). The funders had no role in study design, data collection and analysis, decision to publish, or preparation of the manuscript.

Competing Interests: The authors have declared that no competing interests exist.

* E-mail: men@ciac.ac.cn

## Introduction

Polymers are long chain molecules of random coil structure entangled together in molten state. In most of cases, such highly entangled polymer melt can solidify into a semicrystalline state with layer-like lamellar crystallites stacking together and entangled amorphous chain segments in between. As crystallization of a polymeric system under usual condition does not change a global random coil dimension and does not remove the entanglements, there exist many tie molecules connecting adjacent lamellae [1]. This semicrystalline and hierarchical microstructure feature of semicrystalline polymers makes it very complicated in understanding their structure and properties relationships. Among many other properties, mechanical response of a polymeric system might be the most important one because the majority of polymer materials are used as structural materials such as pipes for transporting oil, gas and water, furniture, containers for food or chemicals, houses for electronic devices and so on. The central topic in the field is thus to understand how the materials behave under mechanical load, usually uniaxial tension. Two major processes have been identified when a semicrystalline polymer is stretched, namely shear yielding and cavitation [2–8]. With respect to the shear yielding, there are two general but distinctly different arguments in the literature. Firstly, the deformation was considered to be accomplished by slips within the lamellae including crystallographic fine slips and intra-lamellar mosaic block slips [9–15]; secondly, stress-induced melting and recrystallization was proposed to be responsible for the deformation process [16–21]. In recent years, investigations based on "true stress−strain experiments" revealed that both processes discussed above may be activated at different strains during tensile deformation [22–26]. Upon stretching, block slippage within the crystalline lamellae took place first, followed by a stress-induced fragmentation and recrystallization starting at certain strain determined by the stability of crystalline blocks and the state of entangled amorphous network [27].

However, cavitation process and the correlation between cavitation and shear yielding have not been understood thoroughly due to the complicated semicrystalline nature of the polymers as being composed of crystalline lamellae and entangled amorphous polymeric chains in between [1]. Butler et al. [28,29] observed that cavitation occurred at the yield point and concluded that cavitation was initiated after the onset of crystal shear, and suggested that crystal shear could favor the generation of cavities. In contrast, Pawlak and Galeski [5,30] proposed that the cavities occurred in the amorphous phase between the lamellae with normal parallel to the tensile direction, and cavitation could be initiated before the yield point. However, Men et al. [31] considered that due to the blocky substructure, the cavities in isotactic polybutene-1 (PB-1) occurred first at the block boundaries of those lamellae with their normal perpendicular to the tensile direction and passed through amorphous phase connecting several lamellae. Bao et al. [32] also observed that the cavities in oriented isotactic polypropylene (iPP) occurred in the lamellae with normal perpendicular to the tensile direction. The cavitation was observed in many drawn polymeric materials such as high-density polyethylene (HDPE) [28,29,33–39], iPP [32,40–48], and PB-1 [7,31,49]. The processes of crystal shearing and cavitation can be activated concomitantly or competitively under tensile loading depending on the structural characteristics of the materials and the experimental conditions. And the microstructure [28,29], stretching temperature [32,48,50] and stretching rate [41,47] are the important parameters to influence the cavitation.

Usually the cavitation is detected by small-angle X-ray scattering (SAXS) technique from tens of nanometers sized voids, seen as a rapid increase of scattering intensity [32–39,45–48,50]. However, cavitation is visible as strain-whitening macroscopically, which means the formation of heterogeneities (commonly cavities) at least at a length scale of the wavelength of visible light, i.e., some hundreds of nanometers. The length scale is not accessible for conventional SAXS. Compared to SAXS, synchrotron ultra-small-angle X-ray scattering (USAXS) has the advantage of giving access to quite small q-values and provides the possibility to study the large-scale structure evolution during the deformation of polymeric materials [31,51,52].

Compared to other polyolefins like PE and PP, PB-1 possesses superior creep resistance and high temperature resistance, and is applied in many fields, e.g., pressurized tanks, tubes, and hot water pipes [53,54]. Aside from its useful physical properties, PB-1 shows interesting polymorphic behavior. PB-1 always forms the metastable tetragonal phase II [55] crystalline modification when crystallized from the melt state under normal conditions, which transforms into the thermodynamically stable form I when kept at room temperature [56–59]. During the transition from crystalline phase II to I a global shrinkage is always observed, which generates many defects in the crystalline lamellae [60]. This becomes even more evident when the blocky substructure of the lamellae after crystallization is considered [61–63]. Blocks exhibit disordered grain boundaries and the shrinkage of the crystalline blocks results in more disordered boundaries between the adjacent blocks. These loose block boundaries can serve as precursors for the cavities during tensile deformation [31].

By means of in-situ USAXS and WAXS measurements, this work is to investigate the characteristics of cavitation in PB-1 as a function of crystalline lamellar thickness and stretching temperature and to study the relationship between cavitation and shear yielding in such systems. As will be shown in the following sections, three different kinds of cavitation behavior in the system can be observed, namely no cavitation, cavitation with reorientation and cavitation without reorientation. It turns out that the later two cases possess different mechanism of generation of voids being within lamellar stacks due to crystalline block boundary separation before yielding for the cavitation with reorientation case and voiding in inter-fibril regions after yielding.

## Materials and Methods

The PB-1 is produced by BASELL Polyolefines with a trade name of PB0110M, whose melt flow rate (MFR) is 0.4 g/10min (190°C/2.16 kg). Pellets of PB-1 were first compression molded into films of about 0.5 mm in thickness at 180°C and held in the molten state for 5 min to erase the processing history. The molten films were transferred rapidly into isothermal water bath at different preset crystallization temperatures ($T_c$) ($T_c$ = 0, 30, 40, 50, 60, 70, 80 and 90°C) and held isothermally for more than 5 hours to complete the crystallization in the samples. The isothermally crystallized PB-1 samples were stored at room temperature for 1 month to allow a complete phase transition from the metastable tetragonal phase II to the stable hexagonal phase I.

SAXS experiments were carried out in a modified Xeuss system of Xenocs SA, France. The system is equipped with a multilayer focused Cu $K_\alpha$ X-ray source (GeniX$^{3D}$ Cu ULD), generated at 50 kV and 0.6 mA. The wavelength of the X-ray radiation was 0.154 nm. The X-ray beam was then collimated with two pairs of Scatterless slits system located 2400 mm apart from each other. Scattering data were recorded with the aid of a semiconductor detector (Pilatus 100K, DECTRIS, Swiss). In case of measuring crystalline lamellar structure, the sample-to-detector distance was 1760 mm, and the effective range of the scattering vector $q$ ($q = \dfrac{4\pi \sin\theta}{\lambda}$, where $2\theta$ is the scattering angle and $\lambda$ is the wavelength) was 0.071–0.796 nm$^{-1}$. Each SAXS pattern obtained in the center of the sample was collected within 30 minutes, background corrected and normalized using the standard procedure. Beside one-dimensional scattering intensity distributions integrated from 2D SAXS patterns, the technique of one-dimensional electron density correlation function analysis has been also used to give detailed structural information of the systems, such as long period ($d_{ac}$), crystalline lamellar thickness ($d_c$) and amorphous thickness ($d_a$). And linear crystallinity ($\Phi_l$) was deduced from the relation $\Phi_l = d_c/d_{ac}$. Furthermore, complimentary USAXS measurements during stretching few samples have also been performed based on this modified Xeuss system. For USAXS measurements, a sample to detector distance of 6453 mm was used yielding an effect $q$ range from 0.016 to 0.272 nm$^{-1}$. USAXS patterns were collected within 180 s for samples showing strain-whitening, and were treated as described above.

DSC measurements were conducted to acquire the melting temperature ($T_m$) and weight fraction crystallinity ($\Phi_w$). A DSC1 Star$^e$ System (Mettler Toledo Instruments, Swiss), which had been calibrated for temperature and melting enthalpy by using indium as a standard, was used during the experiments at a heating rate of 10 K/min. For the calculation of crystallinity, a value for the heat of fusion at 100% crystallinity of $\Delta H_{id}$ = 125 J/g [64] were used.

Structural information of the isothermally crystallized samples is given in Table 1. With the increasing of the crystallization temperature ($T_c$), the melting temperature ($T_m$), long period ($d_{ac}$), lamellar thickness ($d_c$) and amorphous thickness ($d_a$) all increase. It is noticed that the linear crystallinity ($\Phi_l$) shows no crystallization temperature dependency and keeps essentially constant, while weight fraction crystallinity ($\Phi_w$) increased with increasing crystallization temperature due to the secondary crystallization in the process of cooling to room temperature. More specifically, DSC crystallinity ($\Phi_w$) measures a mass fraction of crystallized portion in the sample whereas SAXS linear crystallinity ($\Phi_l$) represents volume fraction of space occupied by the crystalline lamellae within the lamellar stacks. For semicrystalline polymers, SAXS linear crystallinity normally remains unchanged after isothermal crystallization at different temperatures. Similar results on syndiotactic polypropylene and its octane copolymers, poly(-ethylene-co-octene) and poly(ε-caprolactone) has been reported by Strobl et al. that their SAXS linear crystallinity showed no crystallization temperature dependency and keeps essentially constant [65]. It was assumed that the melt was a two-component fluid, composed of crystallizable and non-crystallizable chain parts with essentially fixed fractions, and the potential of a given polymer system to crystallize is limited and well defined over a larger temperature range. However, when the samples were cooled down from the isothermal crystallization temperature to room temperature, secondary crystallization outside of the crystalline lamellar stacks occurred. The crystallites formed during cooling down do not contribute to the SAXS crystallinity due to their irregular arrangement (lack of well defined periodical structure which gives rise the SAXS peak for calculating the linear crystallinity). These secondary crystallites still contribute to the DSC crystallinity as DSC measures the weight fraction of the crystallites in the sample regardless their arrangement.

"Dog bone" tensile bars with dimensions of $10*5*0.5$ mm$^3$ were obtained with the aid of a punch. Uniaxial tensile deformation was carried out at different stretching temperatures ($T_s$) ($T_s = 30$, 45, 60, 75, 90 and 100°C) using a portable tensile testing machine (TST350, Linkam, UK). The cross-head speed during stretching was kept at 20 μm/s. In order to measure the strain of the deformed area located at certain spots on the samples accurately, optical photo images of the samples were employed. The Hencky measure of strain $\varepsilon_H$ is used as a basic quantity of the extension, which is defined as

$$\varepsilon_H = 2 * \ln \frac{d_0}{d} \qquad (1)$$

where $d_0$ and $d$ are the widths of the sample before and after deformation, respectively. It must be mentioned that above equation is based on an assumption that the Poisson ratio of the sample is 0.5 which is not often met by polymeric samples. In addition, cavitation process also affects the accurate determination of the true strain value based on above equation. Nevertheless, the method provides a fast and reliable measure of the approximate strain value.

USAXS measurements were carried out in order to determine the cavitation behavior and long period of crystalline lamellae in the samples as a function of strain. In situ synchrotron USAXS measurements were performed at the synchrotron beamline BW4 at HASYLAB, DESY, Hamburg, Germany. The wavelength of the X-ray radiation was 0.13808 nm, and the distance of the sample-to-detector was 13610 mm. At this distance, the effective range of the scattering vector $q$ was 0.022–0.256 nm$^{-1}$ in horizontal direction and 0.016–0.256 nm$^{-1}$ in vertical direction due to the rectangular shaped beam-stop. USAXS patterns were background corrected and normalized using the standard procedure.

WAXS experiments were carried out in order to determine the crystalline texture changes of the deformed samples as a function of strain. In situ synchrotron WAXS measurements were performed at beamline BL16B at SSRF, Shanghai, China. The wavelength of the X-ray radiation was 0.124 nm, and the distance of the sample-to-detector was 125 mm. Each WAXS diagram obtained in the center of the sample was collected within minutes. The orientation of the lattice plane was calculated using the Hermans orientation equation [66]:

$$S_{hkl} = \frac{(3 < \cos^2 \vartheta_{hkl} > -1)}{2} \qquad (2)$$

where $\vartheta$ is the angle between the normal direction of the corresponding crystallographic plane and the reference axis, and $\langle \cos^2 \vartheta_{hkl} \rangle$ is defined as:

$$< \cos^2 \vartheta_{hkl} > = \frac{\int_0^{\pi/2} I_{hkl}(\vartheta) \cos^2 \vartheta \sin \vartheta d\vartheta}{\int_0^{\pi/2} I_{hkl}(\vartheta) \sin \vartheta d\vartheta} \qquad (3)$$

where $I(\vartheta)$ is the scattering intensity along the angle $\vartheta$. And the cosine of the scattering angle is defined by Polanyi equation [67]:

$$\cos \vartheta_{hkl} = \cos \mu * \cos \theta_{hkl} \qquad (4)$$

where $\mu$ is the azimuthal angle along the Debye circle, and $2\theta$ denotes the Bragg scattering angle. Note that the orientational order parameter in the present case assumes values in the range $-0.5 \leq S \leq 0$. For a perfect orientation of the lattice planes with their normal perpendicular to the stretching direction, the order parameter would be $S = -0.5$, while for an isotropic sample in the randomly oriented state, the order parameter becomes $S = 0$.

## Results and Discussion

### True Stress-Strain Curves

In Figure 1, the true stress-strain curves are presented for PB-1 samples crystallized at 50°C stretched at elevated temperatures and crystallized at different temperatures stretched at 100°C. The true stress-strain curves can be divided into three zones, i.e., pre-yielding region before the yield point at a strain of about 0.1 (O-Y), strain softening region from a strain of about 0.1 to about 0.7 (Y-S), and strain-hardening region up to fracture (S-F) [23]. For PB-1 samples crystallized at 50°C, the yield stress decreased from 17 to 6 MPa with the stretching temperature increasing from 30 to 100°C. In the region O-S, the true stress decreased notably with increasing temperature. However, the samples show much higher stresses in the strain-hardening region when stretched at 90 and 100°C than at 75°C. Corresponding to the mechanical behavior, as indicated in the photographs of the stretched samples, with the increasing of stretching temperature the samples showed firstly a decrease in the extent of strain-whitening reaching the weakest strain-whitening at 75°C followed by an obvious enhancement in strain-whitening at higher temperatures. The change in the behavior of strain-whitening of the samples with increasing stretching temperature was different from the case of PP[47], where the strain-whitening decreased gradually, even disappeared at elevated temperatures. For PB-1 samples stretched at 100°C, the yield stress increased from 6 to 10 MPa and the phenomenon of strain-whitening was enhanced gradually with the increase in crystallization temperature and thus thickness of crystalline lamellae. Interestingly, samples with thinner lamellae showed much higher stress in strain-hardening region than samples with thicker lamellae. It should be mentioned that in some of the photographs of the samples in figure 1 the dark area with sharp boundaries rounded on the ends is due to the background of the metal block heater attached to the stretcher with an ellipsoidal shaped hole for passing through X-ray during in situ measurements. Samples became highly transparent in some cases so that the background showed up.

**Table 1.** Characteristics of Samples.

| $T_c$ (°C) | DSC | | SAXS | | | |
|---|---|---|---|---|---|---|
| | $T_m$ (°C) | $\Phi_w$ (%) | $d_{ac}$ (nm) | $d_c$ (nm) | $d_a$ (nm) | $\Phi_l$ (%) |
| Quenched to 0 | 124.5 | 50.3 | 20.8 | 12.5 | 8.3 | 60.1 |
| 30 | 125.4 | 53.2 | 24.7 | 14.9 | 9.8 | 60.3 |
| 40 | 126.3 | 56.0 | 26.7 | 16.0 | 10.7 | 59.9 |
| 50 | 127.2 | 57.7 | 27.8 | 16.7 | 11.1 | 60.1 |
| 60 | 127.9 | 58.9 | 30.3 | 18.1 | 12.2 | 59.8 |
| 70 | 129.8 | 62.8 | 33.7 | 20.2 | 13.6 | 59.9 |
| 80 | 132.2 | 65.9 | 39.5 | 23.6 | 15.9 | 59.7 |
| 90 | 133.8 | 68.2 | 42.5 | 25.4 | 17.1 | 59.7 |

The mechanical behavior is correlated with the structure variation during stretching. Therefore, it is speculated that the difference in stress-strain curves and the unusual strain-whitening behavior of PB-1 samples may be related to different structure evolution and deformation behavior during stretching. This speculation will be further confirmed by the in-situ USAXS and WAXS analysis in the following sections.

## In-situ USAXS Analysis of PB-1

In-situ USAXS analysis is a powerful tool to characterize the evolution of cavitation and lamellar structure of semicrystalline polymers during stretching. The scattering intensity in USAXS patterns is determined by the difference of the electron densities, including the scattering of cavities in the low $q$ range as well as the lamellar structure between crystalline and amorphous phases. The scattering of cavities is much more intense than that of lamellar structure due to the strongly different electron densities.

In Figure 2, the upper plot shows the selected 2D-USAXS patterns of PB-1 samples crystallized at 50°C which were stretched at different temperatures. The samples before stretching showed an isotropic scattering ring, which was consistent with the scattering from randomly oriented lamellar stacks in a semicrystalline polymer. At the stretching temperature of 30°C, strong horizontal scattering intensity distribution indicating the occurrence of cavitation with plate-like cavities with their normal orientated along stretching direction appeared before the yield point. This strong scattering signal was further enhanced with the increasing of strain and eventually leveled off after reaching certain strain. Around the strain-hardening point, one observed a beginning of redistribution of the scattering intensity from concentrated at horizontal to vertical indicating a change in the orientation of cavities with their normal parallel to perpendicular

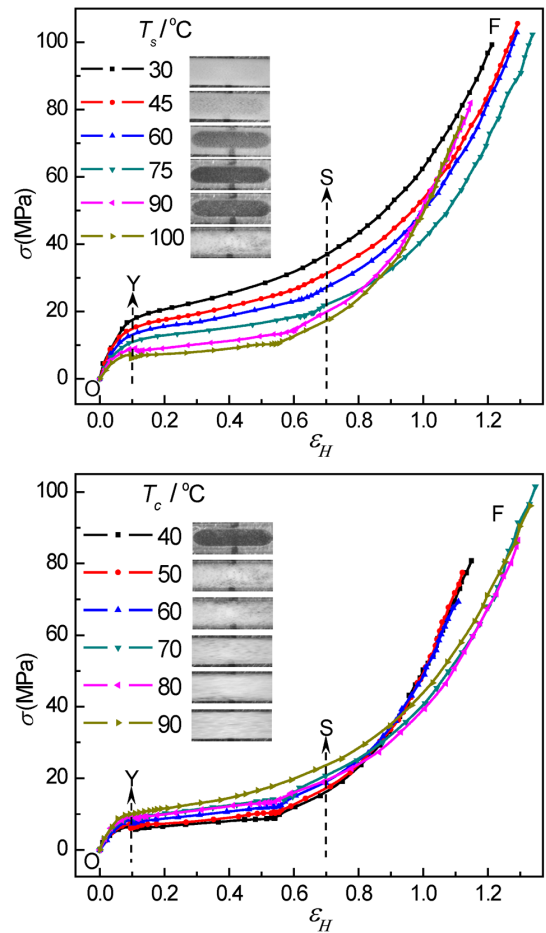

**Figure 1. Selected true stress $\sigma_H$ - strain $\varepsilon_H$ curves of PB-1.** Top: the curves for the samples isothermally crystallized at 50°C stretched at elevated temperatures; bottom: the curves for the samples crystallized at different temperatures stretched at 100°C. Photographs in the inset show the strain-whitening phenomenon of the samples after stretching at different conditions. The characteristic positions within the curves are marked: O refers to the starting point, Y to the yield point, S to the beginning of strain hardening, and F to fracture.
doi:10.1371/journal.pone.0097234.g001

to the stretching direction. This mode of cavitation was termed as "cavitation with reorientation", where the cavities appeared before the yield point. With the increasing of stretching temperature from 30 to 75°C, one observed only a gradual decrease in total scattering intensity indicating a weakened cavitation phenomenon but the sequences of changes in scattering intensity distribution remained the same that all samples showed a behavior of "cavitation with reorientation". However, for stretching temperature of 90 and 100°C, the strain at which cavitation started to appear was obviously beyond the yield point. Moreover, the scattering intensity distribution after cavitation indicated that the cavities occurred with their normal already perpendicular to the stretching direction as soon as they appeared. Increasing in strain led to an increase in the total scattering intensity and an increase in the aspect ratio of the plate-like cavities. Such a behavior was then termed as "cavitation without reorientation". The evolution of cavitation at elevated temperatures was in accord with the strain-whitening phenomenon. This assignment of "cavitation without reorientation" for samples deformed at 90 and 100°C finds supports from the stress-strain behavior shown in

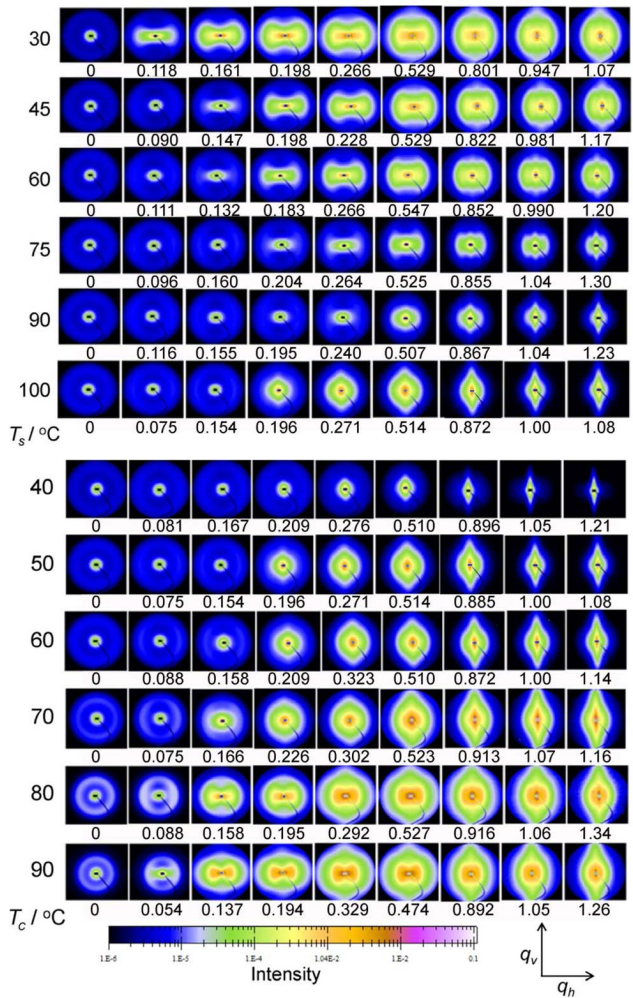

**Figure 2. Selected 2D-USAXS patterns.** The evolution of USAXS patterns of PB-1 taken at different strains. Top: PB-1 samples crystallized at 50°C and stretched at temperatures indicated on the graph. Bottom: PB-1 samples crystallized at different temperatures as indicated and stretched at 100°C. Stretching direction is horizontal. Arrows indicate the directions of integrating intensity.
doi:10.1371/journal.pone.0097234.g002

Figure 1 where in both cases strong strain-hardening was observed as well as a close look at the scattering intensity distribution along and perpendicular to the stretching direction as will be discussed later.

The 2D-USAXS patterns of PB-1 samples isothermally crystallized at different temperatures (with different crystalline lamellar thickness) stretched at 100°C were shown in the bottom. With increasing crystallization temperature, the isotropic scattering rings of original samples became smaller indicating an increase in long period. For the samples crystallized below 70°C with thinner lamellae, the scattering from cavities appeared after yield point and was always concentrated in the vertical direction. The elongation of cavities was along the tensile direction throughout indicating a mode of "cavitation without reorientation". However, for the samples crystallized above 70°C with thicker lamellae, the evolution of cavitation proceeded in accordance with the mode of "cavitation with reorientation".

The evolution of the modes of cavitation of the PB-1 samples as a function of lamellar thickness and stretching temperature was summarized in the sketch in Figure 3. The two parameters both play important roles on the cavitation in PB-1. Based on the difference of the evolution of cavitation mode at different temperatures, the samples can be classified into five groups according to the lamellar thickness. At any temperature investigated, the samples with thickest lamellae (Part I) showed only "cavitation with reorientation" whereas the samples with thinnest lamellae (Part V) presented only "no cavitation" throughout the stretching. For samples with lamellar thickness in between (Part II, III and IV), the evolution of the cavitation mode was more complex. When stretched at elevated temperature, Samples at Part IV experienced from "cavitation with reorientation" to "no cavitation", while samples at Part II did from "cavitation with reorientation" to "cavitation without reorientation". Only the samples at Part III experienced all three modes from "cavitation with reorientation" to "no cavitation" and "cavitation without reorientation" with the stretching temperature increasing.

Clearly, the sequence of the occurrence of cavitation and shear yielding in PB-1 samples depends on lamellar thickness and stretching temperature. For the samples with thicker lamellae or samples stretched at lower temperature, cavities with their normal parallel to the tensile direction appeared prior to the shear yielding and started to reorient with their normal perpendicular to the tensile direction after strain-hardening. However, for the samples with thinner lamellae or samples stretched at higher temperature, the shear yielding occured first and the cavitation could take place or not with increasing strain. In case that cavitation occurred, the cavities were always aligned with their normal perpendicular to the drawing direction.

Figure 4 shows the integrated scattering intensity over whole 2D USAXS images as a measure of the scattering power of PB-1 at different strains. It is found that a pronounced increase in the scattering intensity occurred, showing the appearance of cavities with an electron density largely differing from that of the constituents of the polymeric phase. In addition, the integrated scattering increased with the strain and reached a maximum value in the strain softening region followed by a decrease with further straining. For the PB-1 samples crystallized at 50°C, with the increase of stretching temperature, the scattering intensity first decreased being significantly weakened when stretched at temperatures between 75 and 90°C followed by an increase again at yet higher stretching temperature of 100°C. Moreover, for the PB-1 samples with different lamellar thickness stretched at 100°C, the intergrated scattering intensity increased with the increaseing of lamellar thickness.

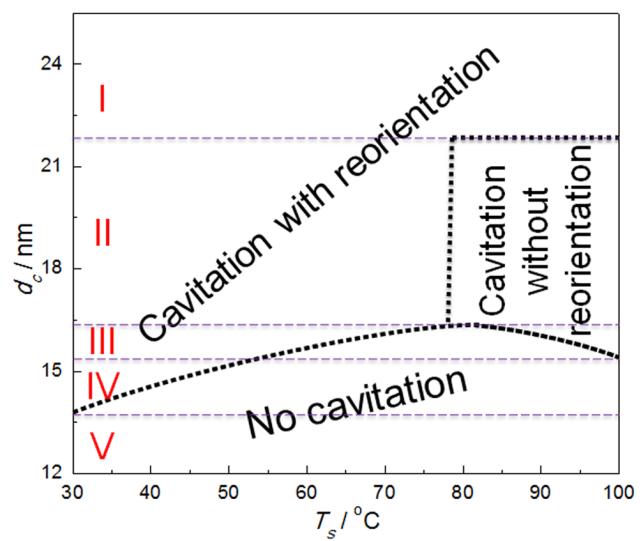

**Figure 3. The diagram of the cavitation modes.** The evolution of cavitation modes of PB-1 as a function of the lamellar thickness and stretching temperature.
doi:10.1371/journal.pone.0097234.g003

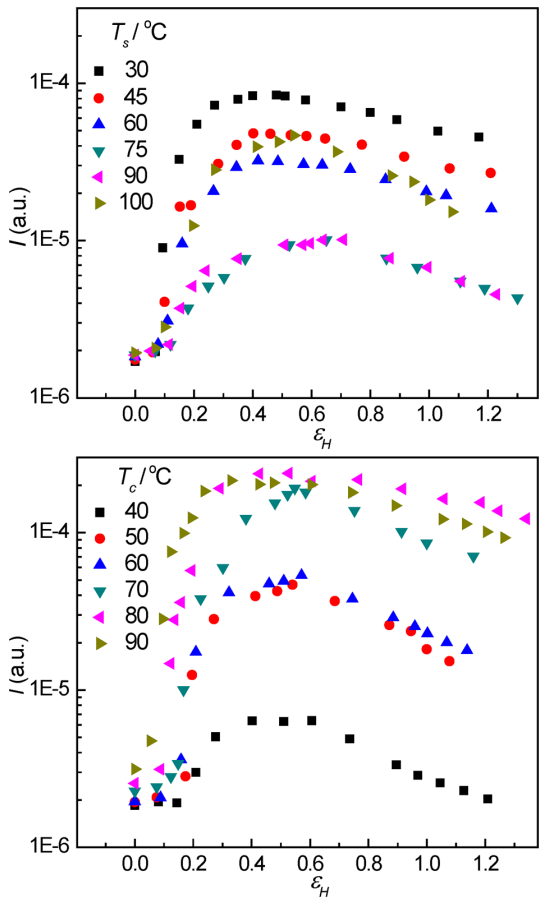

**Figure 4. The integrated scattering intensity.** The evolution of integrated scattering intensity over whole 2D USAXS images as a measure of the scattering power of PB-1 at different strains. Top: the samples crystallized at 50°C and stretched at temperatures indicated. Bottom: the samples crystallized at different temperatures and stretched temperature at 100°C.
doi:10.1371/journal.pone.0097234.g004

Figure 5 shows the evolution of the integrated scattering intensities along (horizontal) and perpendicular (vertical) to the stretching direction and the ratio between them with strains. To a first order approximation, the horizontal and vertical intensities represent the volume fraction of the cavities with their normal parallel and perpendicular to stretching direction, respectively. The change of the ratio is an important evidence to indicate the shape and orientation of the cavities during deformation [50,68]. In Figure 5, the experimental data for samples stretched at 80 and 85°C were collected using our home lab modified Xeuss system, and only the change of ratio is present.

For the PB-1 samples crystallized at 50°C, when stretched at 30°C, the horizontal intensities increased sharply before yield point, then peaked at about a strain of 0.3, and decreased with increasing strain. With increasing stretching temperature till 90°C, the evolution of the horizontal intensities showed similar trend and the maximum intensity was much weaker and the strain of the peaks became slightly larger. When the sample was stretched at

100°C, the scattering intensity from cavities was stronger than when stretched at 75 and 90°C. While the vertical intensities showed the different behavior. Before yield point, the intensities kept almost constant meaning there were hardly cavities with their normal perpendicular to the stretching direction. After yield point, the intensity increased almost linearly with increasing strains. When stretched at 90 and 100°C, the intensity increased first, and then decreased gradually, which differed from the situation at lower stretching temperature. When stretched below 75°C, this ratio was always larger than 1 before the strain-hardening point, indicating that the plate-like cavities were mostly extended in direction perpendicular to stretching direction. During the strain-hardening stage where the ratio was lower than 1 meaning that the orientation of the cavities changed from perpendicular to parallel to stretching direction. However, when stretched above 80°C, the ratio was always smaller than 1 after yield point, indicating that the cavities were always parallel to the stretching direction.

For the PB-1 samples stretched at 100°C, the horizontal intensities showed the similar trend, which increased first followed by a decrease with increasing strain. And with the increase in lamellar thickness, the horizontal intensity increased sharply, indicating that more cavities with their normal parallel to the stretching direction were generated. Meanwhile the vertical intensities increased as well when the crystallization temperature increased from 30 to 70°C, and then decreased with the increase of lamellar thickness, indicating that the amount of cavities with their normal perpendicular to the stretching direction increased first with the increase of lamellar thickness till certain lamella thickness followed by a decrease with further increase in lamellar thickness. The evolution of the ratio between horizontal and vertical scattering intensities further demonstrated the change. The ratio for samples with thinner lamellae was always smaller than 1, meaning that the cavities were along the tensile direction throughout and there was no reorientation of the cavities. However, for samples with thicker lamellae was bigger than 1 at the beginning, reduced with the deformation, and was less than 1 after strain-hardening point, corresponding to the mode of "cavitation with reorientation".

In an effort to elucidate the size of the cavities and lamellar long spacing during deformation process, the one-dimensional scattering intensity distributions in the horizontal and vertical direction are shown in Figure 6 for PB-1 samples crystallized at 50°C and stretched at 30 and 100°C, and the ones crystallized at 90°C and stretched at 100°C. The long period of the lamellar stacks (i.e., the average distance between the mid-planes of two adjacent lamellae measured along the lamellar normal) of the samples was calculated using the plots of Lorentz corrected scattering intensities as a function of scattering vector according to the Equation (5):

$$d_{ac} = \frac{2\pi}{q_{max}} \qquad (5)$$

where $d_{ac}$ is the long period and $q_{max}$ is the peak position.

Although cavities in semicrystalline polymers were often assumed to be ellipsoidal shaped and corresponding dimensions were calculated based on model fitting [37,69,70]. The case for PB-1 is different [31]. For PB-1, during the deformation process, the horizontal scattering intensity distribution showed a maximum at a scattering vector that was smaller than the peak originating from lamellar stacks. This new peak was attributed to the formation of plate-like cavities [31,71]:

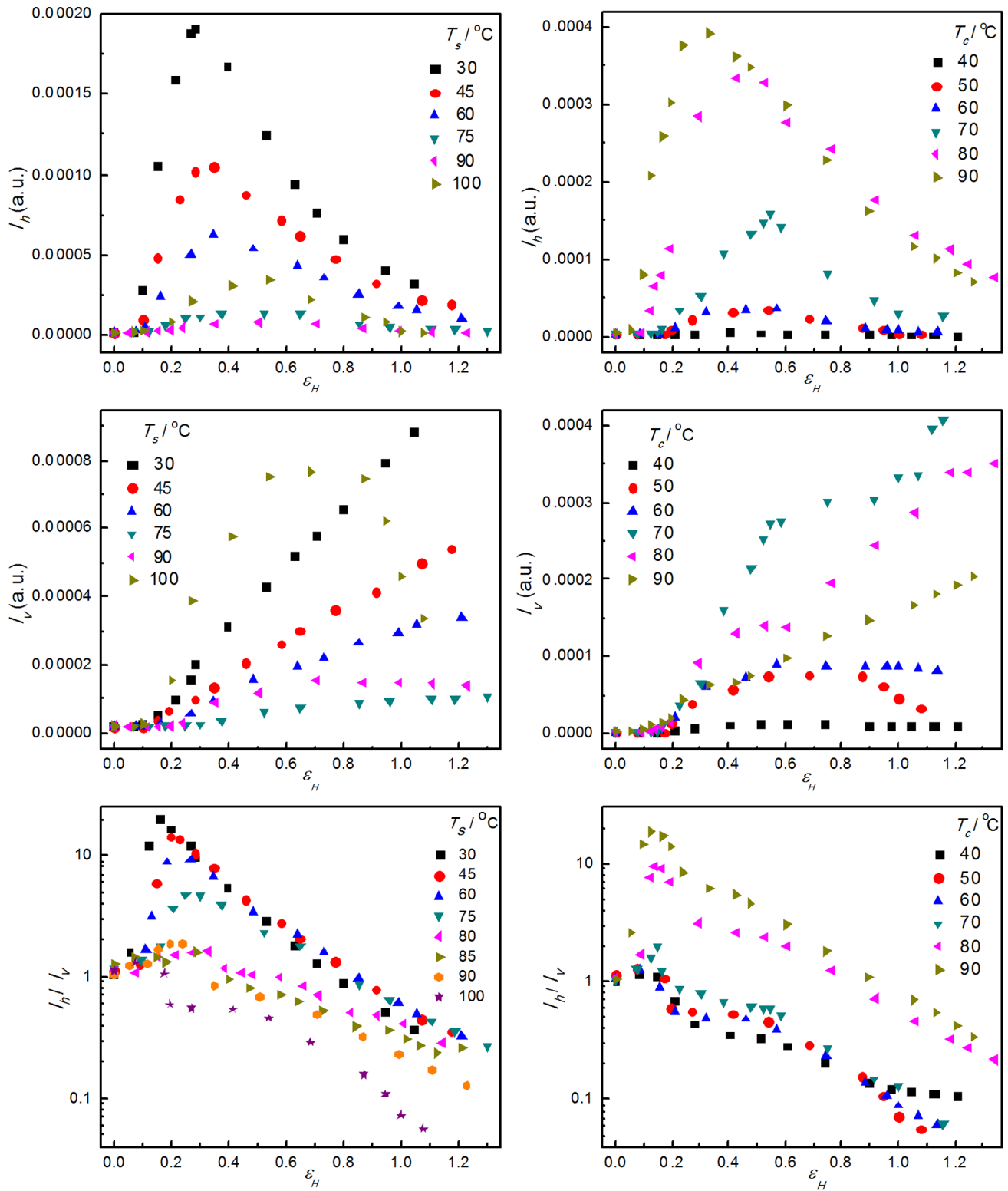

**Figure 5. The horizontal and vertical integrated intensity and the ratio between them.** The evolution of the horizontal (top) and vertical (middle) integrated intensity and the ratio between them (bottom) as a function of strain for the PB-1 samples crystallized at 50°C stretched at different temperatures as indicated (left) and the ones crystallized at different temperatures as indicated and stretched at 100°C (right).
doi:10.1371/journal.pone.0097234.g005

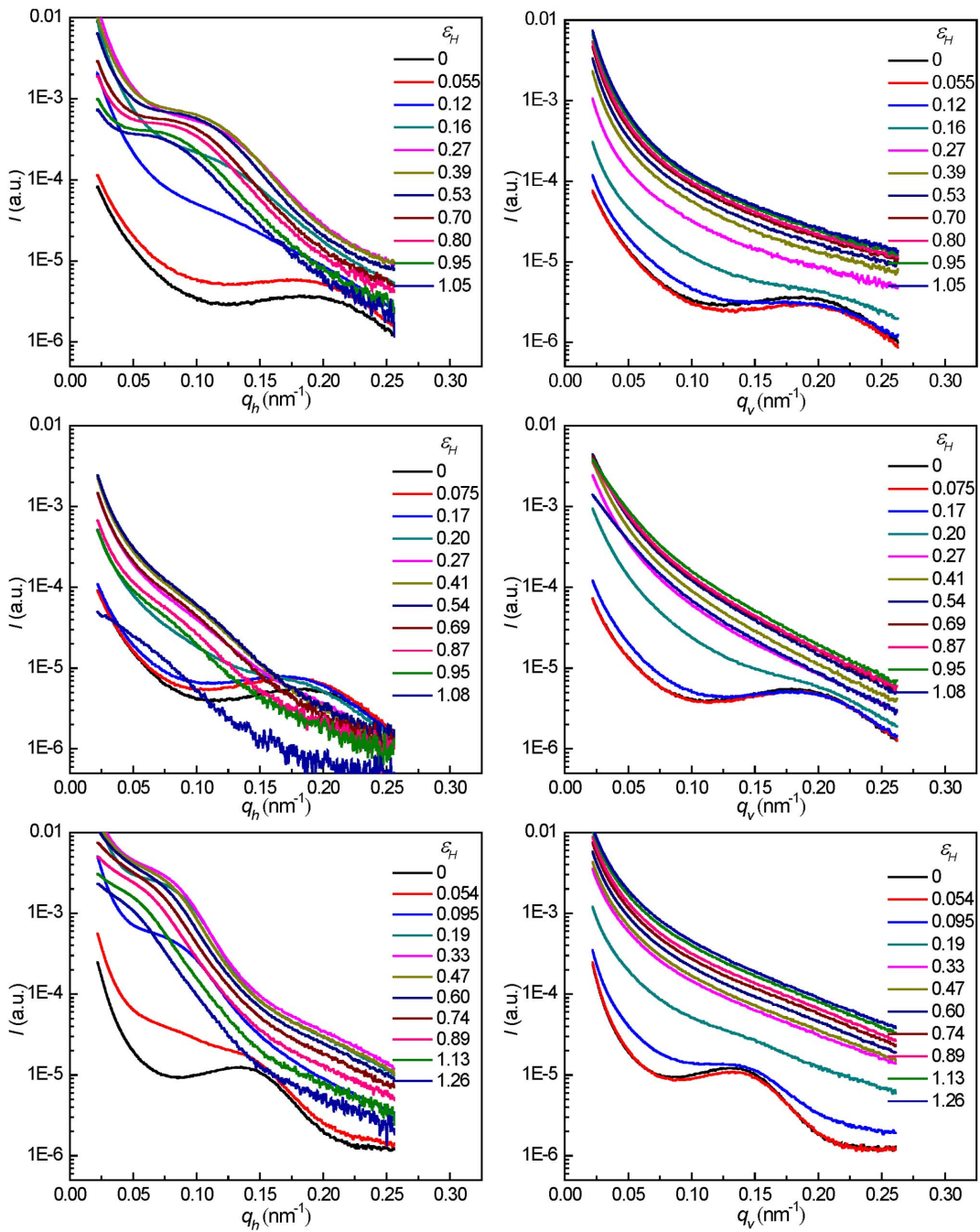

**Figure 6. The 1D scattering intensity distribution profiles along different directions.** Plots of $I$ vs $q_h$ (left) and $I$ vs $q_v$ (right) taken at different strains for PB-1 samples crystallized at 50°C stretched at 30°C (top) and 100°C (middle), and the ones crystallized at 90°C stretched at 100°C (bottom).

$$I(q) = A \frac{2\pi}{q^2} (\Delta\rho)^2 d^2 (\frac{\sin qd/2}{qd/2})^2 \qquad (6)$$

where $A$ is the lateral cross section of the plate, $\Delta\rho$ is the electron density difference between the cavities and its surrounding phase, and $d$ is the thickness of the plate. It is evident from the above Equation (6) that a plot of $I(q)q^2$ vs $q$ yields maxima at $q=0$ and at $q = [(2n+1)/d]\pi$ (where $n = 1, 2, 3,...$). As discussed above, the sample crystallized at 50°C stretched at 100°C experienced the mode of "cavitation without reorientation". Therefore, only the $Iq^2$ vs $q$ plots of the sample crystallized at 50°C stretched at 30°C (top), and the ones crystallized at 90°C stretched at 100°C (bottom) are presented in Figure 7. It was obvious that the intensity maximum shift to smaller $q$ values with the increase of the strain, indicating a continuous opening of the plate-like cavities during stretching. The volume fraction of cavities decreased at large deformation, which was evidenced by the decrease of the scattering intensity. The thickness of cavities along the stretching direction at different strains could be calculated using the maximum scattering intensity position $q_{max}$ according to Equation (7)

$$d = \frac{3\pi}{q_{max}} \qquad (7)$$

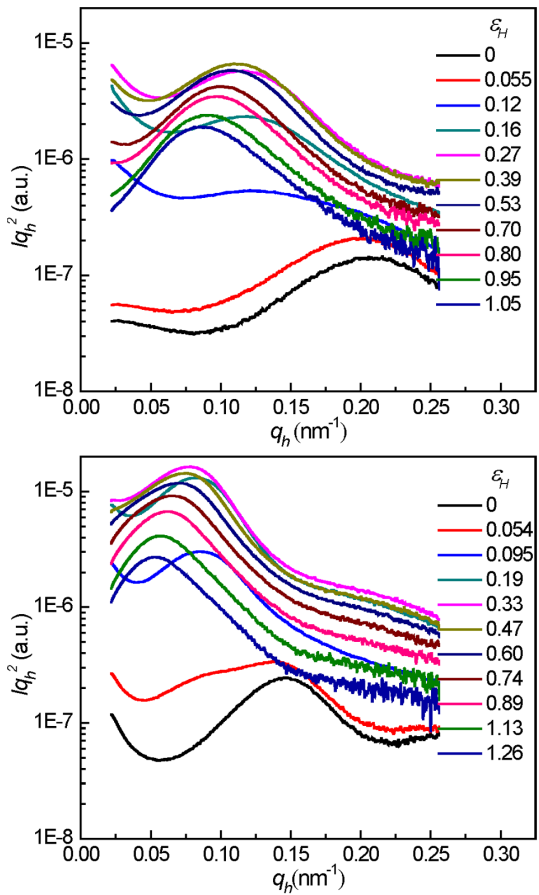

**Figure 7. The 1D scattering intensity distribution profiles with Lorentz correction along stretching direction.** Plots of $Iq_h^2$ vs $q_h$ for PB-1 samples crystallized at 50°C at different strain at 30°C (top), and the ones crystallized at 90°C stretched at 100°C (bottom).
doi:10.1371/journal.pone.0097234.g007

In the top of Figure 8, the thus obtained $d$ values and the long period of lamellar stacks are shown. Due to the strong scattering of cavities, the scattering of lamellar stacks was covered and the long period could only be obtained at low deformations before cavitation. For the mode of "cavitation with reorientation", the evolution of the thickenss of plate-like cavities showed two interesting features. First, for the sample crystallized at 50°C stretched below 75°C, the thickness of cavities was almost same over the whole strain range. And for the samples stretched at 100°C with thicker lamellae, the thickness of the cavities increased with the increase of lamellar thickness. Second, the thickness of cavities increases with the increasing in strain in a linear fashion at first under a smaller slop followed by a change into a larger slop of this linear relationship at the strain-hardening point for all samples.

A linear extrapolation of the cavity thickness to the undeformed state ($\varepsilon = 0$) yielded a nonzero value $d_0$, which was thus assumed to be the smallest stable cavity thickness in the system. Using the value of $d_0$, one was able to calculate the strain of the cavities along its normal by Equation (8)

$$\varepsilon_{H,cavities} = 2 * \ln\left(\frac{d_0}{d}\right) \qquad (8)$$

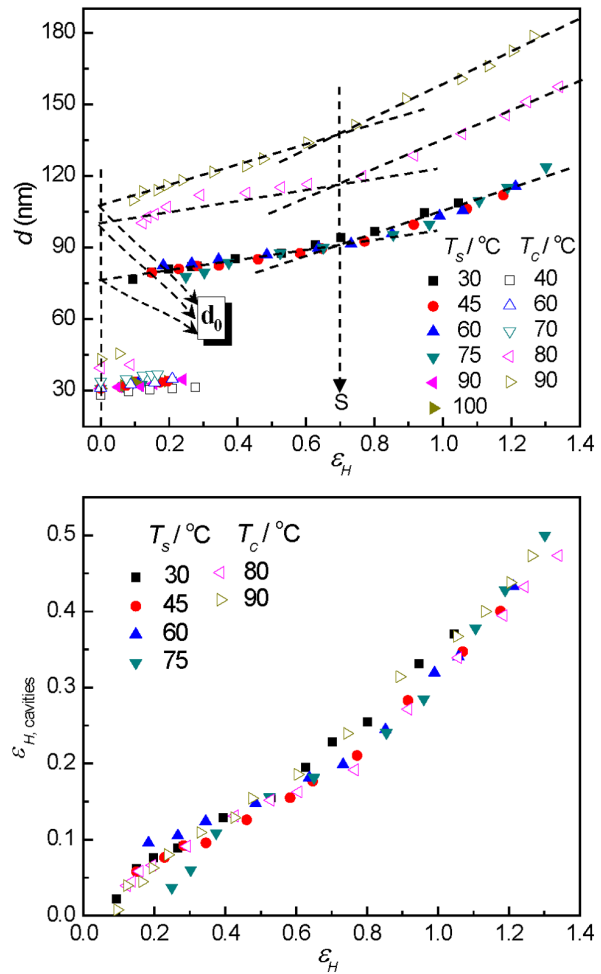

**Figure 8. The evolution of microstructure of lamellae and cavities.** The long period of lamellar stacks and the thickness of the cavities (top) and the master curve of the cavities strain (bottom) along the stretching direction at different strains for PB-1 samples crystallized at 50°C stretched at elevated temperatures (solid symbol) and the ones crystallized at different temperatures stretched at 100°C (hollow symbol). $d_0$ denotes the linear extrapolated thickness of the cavities at zero strain.
doi:10.1371/journal.pone.0097234.g008

The bottom part of Figure 8 showed the master curve of the cavities strain as a function of the macroscopic strain. Interestingly, unlike in the case of strain dependency of thickness where samples with thicker lamellae showed much thicker cavities, the strain of the cavities for all samples followed the same line. In addition, it must be mentioned that the strain of cavities was much smaller than the macroscopic strain, indicating that the deformation was locally inhomogeneous in the system.

In the right part of Figure 6, the vertical intensities at large deformations showed a smooth decay pointing to a wide distribution of the length of cavities with their normal perpendicular to the tensile direction. For the mode of the "cavitation without reorientation", the streak scattering across the beam stop along the vertical direction in SAXS patterns was attributed to cavities with high degree of orientation. Therefore the length of cavities can be estimated by Ruland's method [72]. The length and misorientation of the cavities can be extracted from the data with the help of Equation (9):

$$B_{\mathrm{obs}} = \frac{1}{l} * \frac{2\pi}{q} + B_{\Phi} \tag{9}$$

where $B_{\mathrm{obs}}$ denotes the integral breadth (peak area/peak height), $l$ denotes the length of cavities and $B_{\Phi}$ denotes the misorientation of cavities. A series of azimuthal scans at different values of scattering vectors $q$ along the vertical direction were performed and fitted with a Lorentz function. $B_{\mathrm{obs}}$ is plotted as a function of $1/q$. The length of the cavities can be obtained from the slope of the data and the misorientation is given by the intercept. Figure 9 shows the length and the misorientation of the cavities as a function of strain for PB-1 samples crystallized at 40, 50, 60 and 70°C stretched at 100°C (for the mode of "cavitation without reorientation"). Obviously, the length of cavities increased with strain and always lied in the wavelength range of visible light, which contributed to the strain-whitening phenomenon. The misorientation reflects the degree of orientation of cavities. With increasing strain, the misorientation decreased gradually, that is, the cavities aligned more and more along the tensile direction. For sample crystallized at 40°C, longer cavities were found than other samples. This phenomenon can be attributed to the fact that thinner lamellae in the sample crystallized at 40°C were easily broken down producing less number of cavities. Therefore, there was more space for the individual cavity, and the longer cavities were generated.

## In-situ WAXS Analysis of PB-1

WAXS experiments were performed in order to explore relationship between changes of crystalline texture by determining the degree of orientation of the crystallites induced by uniaxial drawing and the mechanical behavior and strain-whitening phenomenon. The discussion was based on analysis of the azimuthal intensity distributions along the diffraction circles. Selected WAXS patterns for PB-1 samples crystallized at 50°C stretched at 30 and 100°C, and the ones crystallized at 90°C stretched at 100°C at different strains were presented in Figure 10. The evolution of the reflection of inner ring (110) and outer ring (221+220) was different at different conditions. The evolution of WAXS patterns in the top and bottom was similar. The diffraction intensity of (110) plane was a little stronger at the two shoulders of the vertical direction before strain-hardening point and concentrated on the vertical direction after strain-hardening, accompanying with the mode of "cavitation with reorientation". However, the diffraction intensity in the middle WAXS patterns was focused on the vertical direction from the middle strain and oriented more with followed strain corresponding to the mode of "cavitation without reorientation". Moreover, for large strains, the eight-point pattern of outer ring (221+220) was formed for samples where reorientation of cavities occurred, but six-point pattern was observed for the sample where reorientation of cavitation was not observed. This difference in the occurrence of (220) diffraction spots along stretching direction in case of cavitation with reorientation and absence of such (220) diffraction spots along stretching direction in case of cavitation without reorientation provides significant information in understanding the actual cavitation process in the system at different conditions. The appearance of the (220) diffraction spots along the stretching direction suggested that there are significant portion of crystalline lamellae with their normal perpendicular to the stretching direction preserved after even large macroscopic deformation suggesting that the cavitation process under such condition proceeds indeed within those lamellae stack with their normal perpendicular to the stretching direction. The results also validate our treatment of the calculation of the thickness of the plate-like cavities during stretching. It must also be mentioned that the lack of corresponding (110) diffraction spots along stretching direction is due to the uneven distribution of the lamellar crystallites with their normal perpendicular to the stretching direction. Only if when the system possesses a uniaxial symmetry with fiber axis perpendicular to the stretching direction can such (110) diffraction spots along stretching direction be safely observed together with the corresponding (220) diffraction spots. Clearly, current case does not fulfill such condition. In case cavitation without reorientation occurred, a strong uniaxial orientation of the crystalline lamellar normal along stretching direction is observed indicating a different mechanism of cavitation in the system.

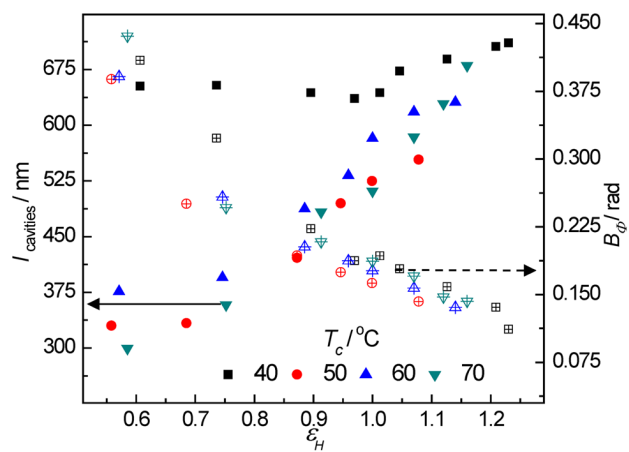

**Figure 9. The length and misorientation of cavities.** The length (full symbols) and misorientation (cross symbols) of cavities as a function of strain for PB-1 samples crystallized at 40, 50, 60 and 70°C stretched at 100°C.
doi:10.1371/journal.pone.0097234.g009

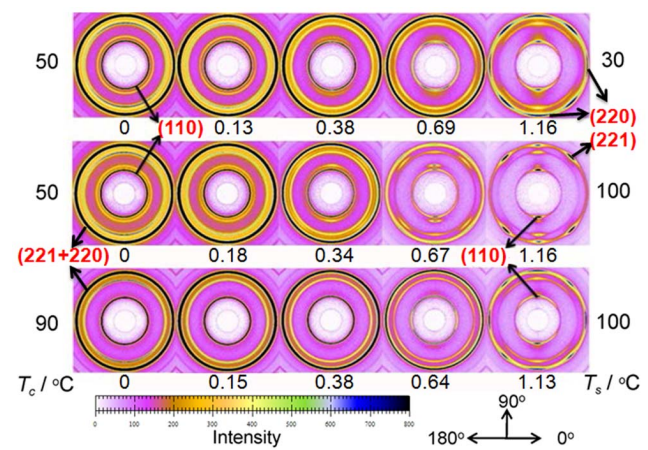

**Figure 10. Selected 2D-WAXD patterns.** The 2D-WAXD patterns taken at different strains as indicated on the graph for PB-1 samples crystallized at 50°C stretched at 30 (top) and 100°C (middle), and the ones crystallized at 90°C stretched at 100°C (bottom). Stretching direction is horizontal. Arrows indicate the azimuthal angle of intensity distribution.
doi:10.1371/journal.pone.0097234.g010

Figure 11 depicts the azimuthal scans of the intensity distribution of the 110-reflection. For the sake of clarity, only the data from 0 to 180° are presented. For PB-1 samples crystallized at 50°C, when being stretched at 30°C, the diffraction intensity in the horizontal direction kept almost constant at lower strain indicating that the lamellae with normal perpendicular to the tensile direction partly broke into the blocks and the chains in the blocks kept the orientation, where the plate-like cavities perpendicular to the tensile direction were generated between those blocks. With increasing in strain, all polymer chains tended to be gradually oriented along the stretching direction contributing to a reduction in the diffraction intensity and resulting in reorientation of cavities. At the strain-hardening, the intensity at an oblique angle was a little stronger due to the stability of blocks. With the further increase in strain, the diffraction intensity at horizontal direction decreased strongly while the vertical one increased gradually due to the formation of the micro-fibrils composed of newly formed lamellar stacks with normal along the tensile direction. However, at 100°C, the evolution of the intensity was different. The diffraction intensity decreased slightly at all directions after yield point due to the changes in scattering volume during stretching, where the ellipsoid cavities along the tensile direction began to be generated. During strain-hardening stage, the horizontal intensity reduced further, while the vertical intensity increased gradually due to the preferential orientation of the newly formed crystalline lamellae. The evolution of diffraction intensity for PB-1 samples crystallized at 90°C deformed at 100°C was similar to the one for the PB-1 samples crystallized at 50°C deformed at 30°C showing a deformation temperature-crystalline lamellar thickness equivalence behavior.

Figure 12 shows the degree of orientation of crystallographic (110) plane $S_{110}$ at different strains. The original samples showed slight orientation. The degree of orientation weakened before yield point. Before the strain-hardening point, the degree of orientation changed slightly for the samples stretched at lower temperature or for the samples with thicker lamellae. However, the degree of orientation increased quickly for samples with thinner lamellae stretched at higher temperature. These results indicated that the lamellar crystallites experienced different deformation processes.

## Deformation mechanism

As was discussed in the introduction section, plastic deformation in semi-crystalline polymers without extensive cavitation during tensile deformation is accomplished mainly by intra-lamellar block slips at small strains. The result of such block slips yields a successive decrease of the orientational order parameter of crystallographic (110) plane indicating a preferential orientation of polymer chains in the crystalline phase towards stretching direction [22]. On the other hand, if inter-lamellar slips were activated, the corresponding orientational order parameter of crystallographic (110) plane in this system would increase due to the orientation of lamellar normal perpendicular to the stretching direction [22]. Data present in Figure 12 show both cases occurring at different conditions. At low stretching temperature or when the lamellar thickness is large, inter-lamellar slips take place resulting in an increase in orientational order parameter at the early stage of deformation. For samples with thinner lamellae and stretched at higher temperatures, one observed at the most beginning of stretching a rather constant orientational order parameter indicating a competition between two slip processes mentioned above followed by the occurrence of dominant intra-lamellar block slips giving a decrease in orientational order parameter. The results can be understood as follows: at low temperatures or for samples with thicker lamellae block slips cannot be activated before cavitation due to the high stress level needed to trigger block slips [6]. Indeed, Men *et al* [62,63] demonstrated that a relaxation process of intra-lamellar block motions in HDPE and syndiotactic polypropylene could be activated at high temperature or low frequency side of the dynamic glass process in dynamic mechanical analysis measurements, and retarded with thickening of lamellae. Similar relaxation process is believed to exist in PB-1 as clear electron microscopy evidence for such motions in PB-1 has been reported [73]. By comparing results present in Figure 5 and 12, one could recognize a corresponding relationship between the ratio of the horizontal and vertical USAXS intensity and the degree of orientation of 110-reflection obtained from WAXS measurements. When the samples with thicker lamellae were stretched at lower temperatures, the ratio was big at low strains, reduced to 1 at the strain-hardening point, and decreased further, while the degree of orientation changed slightly at low strains and increased obviously at the strain-hardening point. However, when the samples with thinner lamellae were stretched at higher temperatures, the ratio was always smaller than 1 and the degree of orientation increased remarkably after yield point. The relationship indicates that the orientation of the crystallites influences the evolution mode of cavitation.

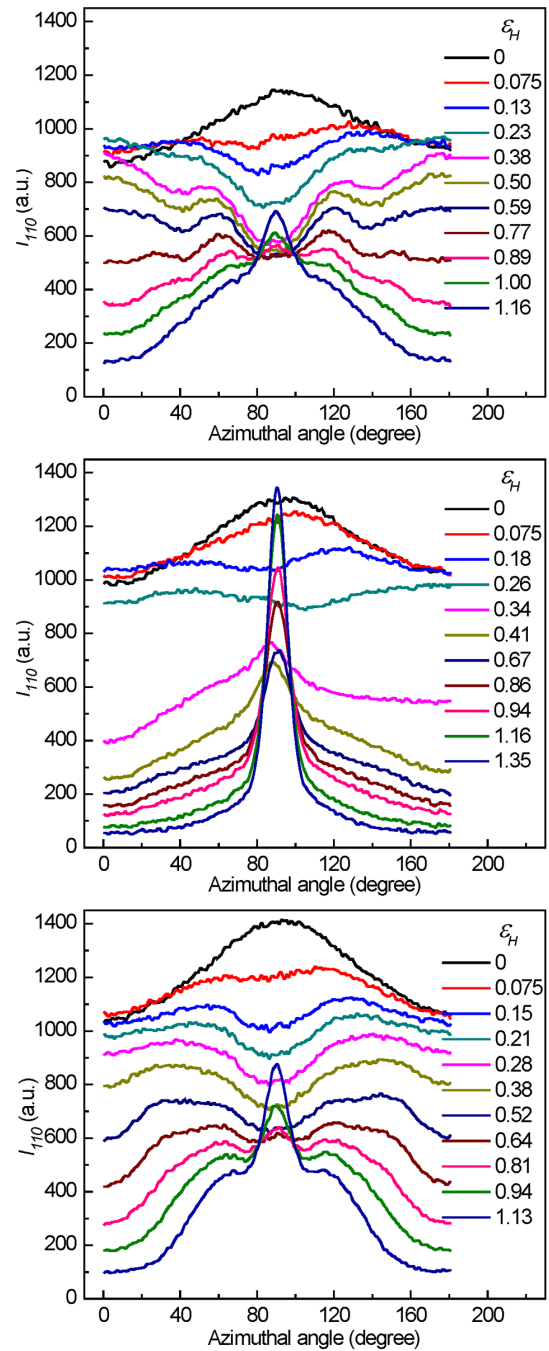

**Figure 11. Azimuthal intensity distribution of 110-reflection.** The evolution of azimuthal intensity distribution of 110-reflection as a function of strain measured for PB-1 samples crystallized at 50°C stretched at 30 (top) and 100°C (middle) and the ones crystallized at 90°C stretched at 100°C (bottom).
doi:10.1371/journal.pone.0097234.g011

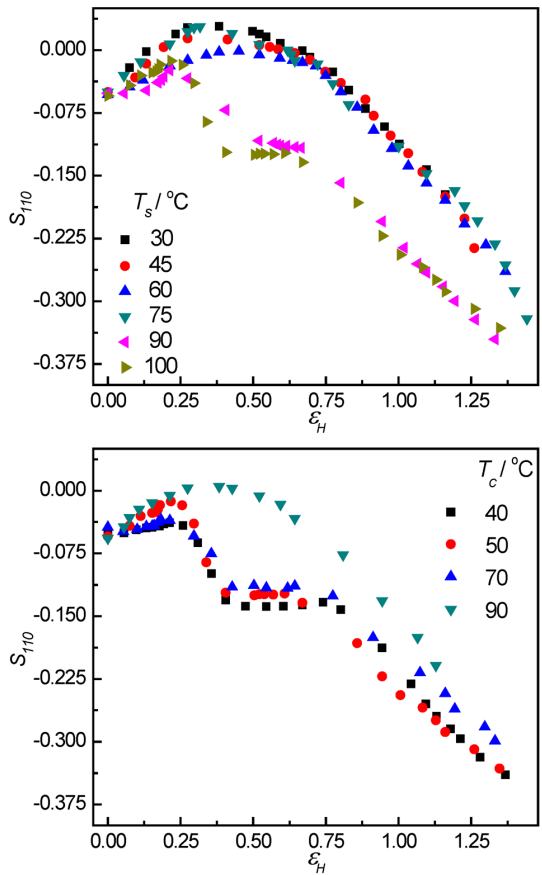

**Figure 12. The degree of orientation of 110-reflection.** The evolution of the degree of orientation of 110-reflection ($S_{110}$) associated with the 110-reflection as a function of strain for PB-1 samples crystallized at 50°C stretched at elevated temperatures (top) and the ones crystallized at different temperatures stretched at 100°C (bottom). doi:10.1371/journal.pone.0097234.g012

A schematic representation of the evolution of cavitation in PB-1 during uniaxial deformation as a function of stretching temperature and lamellar thickness is shown in Figure 13. For the samples stretched at lower temperatures and samples with thicker lamellae, they experienced the mode of "cavitation with reorientation", where the block motions were restrained and the yield stress was higher than the cavitation stress separating crystalline blocks within lamellae. Therefore, the block boundaries within crystalline lamellae with normal perpendicular to the tensile direction opened in width and formed the plate-like cavities passing through amorphous phase connecting several lamellae before yield point. In such case, the block slips occurred only to a limited extent. After yield point, a collective activation of interlamellar and block slips occurred. The plate-like cavities perpendicular to the tensile direction became bigger. Here the polymer chains were perpendicular to the stretching direction. During strain-hardening stage, the part of the polymer chains oriented to the stretching direction resulting in a reorientation of the cavities with their normal perpendicular to the stretching direction. As was discussed before on the WAXD patterns, there was still part of crystalline lamellae with their normal perpendicular to the stretching direction left even after large macroscopic deformation. This evidence, together with those reported in-situ AFM investigations [7], reinforced our assignment of this kind of cavitation to the opening of the block boundaries with those lamellar stacks with normal perpendicular to the stretching direction. Moreover, it can also be understood rather intuitively as follows: tearing a highly oriented plastic band is much easier in its orientation direction rather than the perpendicular one meaning that separation of crystalline block boundaries is much easier than separating adjacent lamellar crystallites being linked by many entangled tie chains.

For the samples with thinner lamellae stretched at higher temperatures, they experienced the mode of "cavitation without reorientation". Here the contact forces between the blocks in the lamellae were weakened intensely at higher temperature. This could arise from a softening of the lateral surfaces of the crystal blocks, so that grain boundaries existing at low temperature were transformed into fluid-like interlock regions, or a mobilization and growth of the latter ones [60]. Then block slips occurred easily while cavities were restrained before yield point. After yield point, the block slips enhanced further, accompanying with the orientation of lamellae and polymer chains along the stretching direction, where ellipsoidal shaped cavities were gradually generated between stacks of lamellae with normal along the stretching direction. Generation of such cavities has its origin linked to the crystalline phase transitions during stretching the PB-1 samples with stable form I. When being stretched at elevated temperatures yet much lower than the final melting temperature, the stable form I crystallites within the PB-1 samples undergo a stress induced melting and recrystallization process into metastable form II from certain strain after the activation of lamellar block slips (yielding)[21]. Such transition continues with further stretching but also accompanied by a gradual phase transition from the newly formed form II to stable form I due to the stress acted on the sample along their normal direction. The form II to form I transition results in a decrease in the diameter of the fibrils composed of newly developed stacked form II lamellae. Due to the global rigidity of the specimen, cavities along the long axis of fibrils form.

## Conclusion

The evolution of the cavitation in stable phase I PB-1 samples of different lamellar thickness subjected to uniaxial tensile deformation at elevated temperatures from 30 to 100°C was examined as a function of the imposed strains by means of combined in-situ synchrotron USAXS and WAXS techniques. PB-1 shows a peculiar temperature dependent strain-whitening behavior where the extent of strain-whitening weakens with the increasing of stretching temperature reaching a minima value followed by an increase at higher stretching temperatures. Correspondingly, a stronger strain-hardening phenomenon was observed at higher stretching temperature. With respect to the cavitation process, three characteristic features can be distinguished during tensile deformation of PB-1 samples with different crystalline lamellar thicknesses and at different stretching temperatures. First, there was "no cavitation" for the quenched sample with the thinnest lamellae, and only shear yielding occurred. Second, at low stretching temperature and for samples with thicker lamellae, cavitation occurred in a mode of "cavitation with reorientation", where plate-like cavities with their normal along stretching direction occurred in those lamellae with their normal perpendicular to the stretching direction before yield point and reoriented after strain-hardening. Third, a cavitation mode of "cavitation without reorientation" took place when the samples possessed thinner lamellae and were stretched at elevated temperatures. In this case, ellipsoidal shaped cavities with their long axis preferentially oriented along stretching direction were generated after yield point, which continuously improve their orientation at the later stage of deformation.

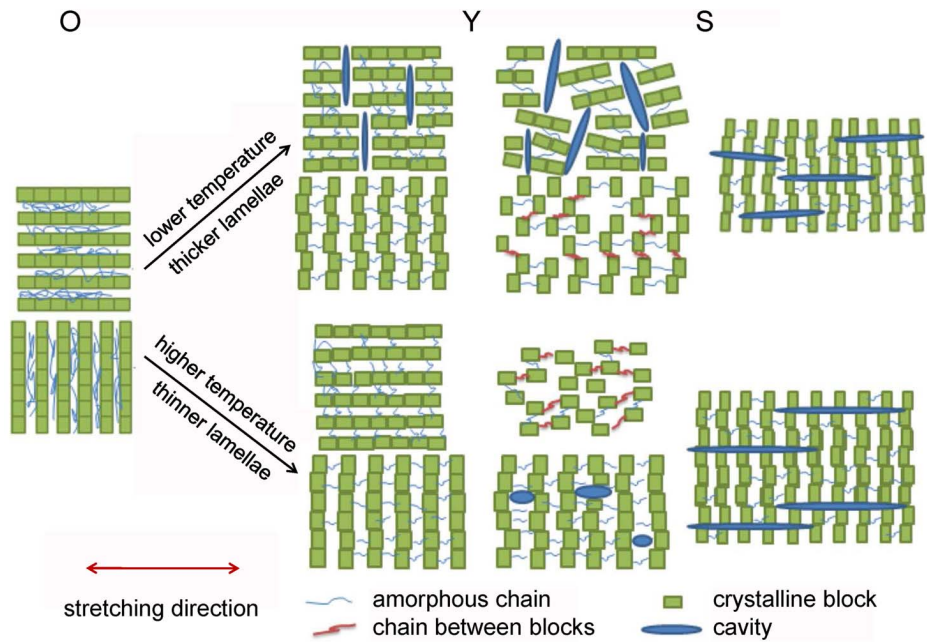

**Figure 13. The cavitation model.** Schematic model of the evolution of cavitation in PB-1 during stretching. Red arrow indicates that stretching direction is horizontal.
doi:10.1371/journal.pone.0097234.g013

## Acknowledgments

We thank Dr. Xiuhong Li at SSRF and Dr. Jan Perlich at HASYLAB for their assistances during synchrotron X-ray scattering experiments.

## Author Contributions

Conceived and designed the experiments: YM YW. Performed the experiments: YW ZJ LF YL. Analyzed the data: YW YM ZJ. Wrote the paper: YW YM ZJ.